\documentclass[conference]{IEEEtran}
\IEEEoverridecommandlockouts
\usepackage{cite}
\usepackage{amsmath,amssymb,amsfonts}
\usepackage{algorithm} 
\usepackage{algpseudocode} 
\usepackage{graphicx}
\usepackage{textcomp}
\usepackage{xcolor}
\usepackage{tikz}
\usetikzlibrary{arrows}

\def\BibTeX{{\rm B\kern-.05em{\sc i\kern-.025em b}\kern-.08em
    T\kern-.1667em\lower.7ex\hbox{E}\kern-.125emX}}
\begin{document}

\title{Two Social Concepts in Virtual Communities: Trust and Reputation\\}

\author{\IEEEauthorblockN{Yusuf Samil Ezer}
\IEEEauthorblockA{\textit{Intelligent Systems} \\
\textit{Christian-Albrechts-Universität zu Kiel}\\
Kiel, Germany \\
stu224931@mail.uni-kiel.de}
}

\maketitle

\begin{abstract}
Our social interactions mainly depend on the social phenomenon called trust. We evaluate our trust in our peer to decide whether to start an interaction or not. When our information about the peer is not sufficient, we use the knowledge of others. This knowledge can also be referred to as the reputation of the peer in the community. Like real-life communities, trust and reputation play a key role in virtual communities, too. These two notions help us overcome the complex interactions between agents in virtual communities. In previous studies regarding this topic, the social aspect of trust and reputation is partly ignored. In this paper, we will review an article which we accept as a starting point and compare it with another article that provides a more advanced model. Additionally, a new trust model which is mainly based on sociological notions will also be introduced.
\end{abstract}


\section{Introduction}
One of the most important notions affecting our everyday social life is trust. Most of the interactions in a society depend on it and it is a very fundamental concept in our daily life \cite{b1}. For example, when we are shopping online, first we evaluate the trustworthiness of the website to be sure if shopping is secure. Trustworthiness in this context is based on our previous knowledge about the website. If we do not have enough knowledge to evaluate the trustworthiness, then we check the information of others, or the reputation of the website. We also use trust and reputation for example to choose the product that we want to buy on the website. The comments for the product, our trust in the brand, or whether we used the product before play an important role in the decision-making process.

Trust not only affects our social life but it also has repercussions in a variety of fields such as economy and politics. As trust provides us with the internal security in an uncertain environment in which we lack the necessary information \cite{b2}, it is an essential notion in our lives. Trust also plays an essential role in virtual communities such as self-organized systems. In these communities, healthy communication between agents is established with the help of trust and reputation. So it is required to transfer these concepts from social life to the field of informatics. However, it is not that easy to define and represent them mathematically for computation. 

The first problem is that, because trust is an abstract concept, it does not have a universal definition that can be used in all research areas. The researchers in the social sciences have proposed different approaches to define trust in a social sense. Our first goal is to find a suitable definition based on the literature in this field. Secondly, we have to formulate this definition mathematically in a trust model which determines the trustworthiness of the other side. Until today, various trust models have been developed by researchers for the computational area. They have used different definitions of trust and reputation and improved different strategies for their model. For the purposes of this paper, we have chosen two different fundamental studies in this area. The first one, which we will accept as the base article for our research, is the system of Abdul-Rahman et al. \cite{b2}. They approach trust and reputation from a social point of view and provide a model in this sense. We have also used this method in our trust model. The second study which we will accept as the reference point is the article of Tomforde et al. \cite{b3}. They have developed a very advanced model which inspired us when we are developing our model. For convenience, the article ``Supporting Trust in Virtual Communities'' \cite{b2} is referred as the ``base article'', the article ``Representation of Trust and Reputation in Self-Managed Computing Systems'' \cite{b3} is referred as the ``reference article'' for the remainder of this paper.

The outline of this paper is as follows: In section II, definitions for trust, computational trust, and reputation are given to provide a common perspective when evaluating these concepts. Section III discusses the two articles that we mentioned above. In this section, we have analyzed the base article in detail and compared it with the reference article. In section IV, we have introduced our trust model. Lastly, section V concludes the paper.

\section{Trust and Reputation}
Trust and reputation are two nested notions that are strongly related to each other. We can understand them better with the help of an analogy. Data and information have also a very comparable relationship to these concepts. Data is a pure fact that does not say anything without an interpretation. But information is the interpreted and processed result of data or collection of data. Very similar to this relation, reputation is the knowledge about an entity acquired in the past and does not give us an opinion without an interpretation and trust gives us the ability to interpret this data. To be more precise, reputation is the knowledge that we had in the past and trust is the result of this knowledge to make more reliable decisions.

After this basic explanation of trust and reputation, we can define them more elaborately and take a look at how they are defined in the literature.

\subsection{Definition of Trust}
Trust can be assessed from different perspectives. For example, in \cite{b4}, Marsh discusses the concept of trust in a variety of ways such as sociological, biological, etc. But our focus will be on the sociological side of it. Since trust arises naturally in social environments and is a feature of collective units like groups or collectivities \cite{b5}, it cannot be conceived in the sense of separate individuals. Gambetta's definition below \cite{b6} of trust is commonly used in literature and we will also it as a starting point in this article:
\begin{quote}
``... trust (or, symmetrically, distrust) is a particular level of the subjective probability with which an agent assesses that another agent or group of agents will perform a particular action, both before he can monitor such action (or independently of his capacity ever to be able to monitor it) and in a context in which it affects his own action''
\end{quote}
Based on this explanation, we can say trust is the subjective prediction of a trustor that a trustee behaves in a certain way. Therefore, there is no universal method to evaluate the trustworthiness of a trustee because the evaluation process depends on the individual perception of the trustor. However, we will take into account the characteristics of trust which are introduced in the base article. This should help us to have a common perspective when defining trust.

\subsection{Computational Trust}
Over the last few decades, the requirements for computation in the area of computer science has seen a gigantic growth. To satisfy this need, computations have substantially moved from a single computer to the distributed systems. However, some further techniques, policies, and processes are needed to provide healthy communication between peers in a distributed system. Especially for choosing a ``good'' peer, we must have a reasonably good decision system.

In decentralized systems, which resemble distributed systems, there is no centralized administration. So the participants of the system cannot identify the other side by requesting a certificate from an authority such as a ``Certificate authority''. Therefore, the agents in the system have to contact each other to make acquaintance with the other side. The purpose of the computational trust is to grant a valid decision by choosing a peer.

Computational trust tries to imitate the real world trust that we have in our social life. Until today, the researchers in this field proposed various trust models. In these models, trust is generally represented as a normalized numerical value between 0 and 1 \cite{b7}. Some other models use different representations for example a binary metric that only takes 0 or 1 as a value. Trust models usually work in the following way:
\begin{enumerate}
\item Input (knowledge about the agent) is given to the trust model (The input can be based on previous interactions or the reputation of the agent in the community).
\item Based on that input trust is calculated.
\item Finally, a decision is made about the trustworthiness of the agent.
\end{enumerate}

\subsection{Reputation}
Similar to trust, reputation is also a concept that comes from daily social life. Therefore, we need to define reputation in a social sense first. Reputation in social groups is the evaluation of a participant by other participants. The behaviors of the participant in interactions shape its reputation in a good or bad way. Although sometimes we have enough information about the participant to evaluate its trustworthiness, our information is usually not sufficient for an evaluation. If we had complete information, then of course we would not need trust anymore \cite{b2} because then, we could predict what the entity will do. As this is not possible, we can use reputation to cope with these uncertainties in daily life by choosing reliable people \cite{b8}.

\section{Related Work and Comparison}
In the area of computational trust and reputation, the researchers developed and proposed several models. One of the main goals of this paper is to review the base article, which we will accept as a starting point, analyze its applicability, and elaborate on it. In this section, we will also compare the base article with the reference article. The both articles are mentioned in section I.

\subsection{Abdul-Rahman and Hailes}

First, we present the base article and its applicability.

\subsubsection{Definition of the Trust and Reputation}
The base article mainly tries to implement characteristics of real-world trust in the virtual medium. For that purpose, they have referred to the social sciences and defined trust and reputation in a social sense. Thereafter they have listed some characteristics of trust in social life which are also important in the virtual-implementation. All of these listed characteristics of trust were applied in their trust model. Since the model is essentially adjusted according to external and internal information, the notion of reputation has especially great significance in this model.
\subsubsection{The Trust Model}
The trust model is centered on the experiences of the agent itself and the recommendations from other agents. It combines these two information resources with a specific method and gets a final trust degree. In this section, we will give a short explanation of the trust system.

In the model, there are four trust degrees ($td$) which are ``\textit{very trustworthy (vt)}'', ``\textit{trustworthy (t)}'', ``\textit{untrustworthy (ut)}'' and  ``\textit{very untrustworthy (vu)}''. Every evaluation of the trustworthiness of an agent should match at the end to one of the four degrees. But they are not general decisions, rather every trust degree should be associated with a context, because trust is context-dependent according to the article. For example, we can say ``I trust Mr. X for car repairing.'' but ``I don't trust Mr. X for housekeeping''. So the trust degree $td$ of an agent $a$ about an agent $b$ should have context $c$. This is called in the model \textit{direct trust} because the agent $a$ evaluates the trustworthiness of $b$ after a direct experience.
\begin{equation*}
t(b,c,td)
\end{equation*}

The second usage of trust is the evaluation of the trustworthiness of recommenders, which is called \textit{recommender trust} $(rt)$. An agent $a$ evaluates the trustworthiness of a recommender agent $b$  ($rtd$)  in a context $c$. With that, the agent $a$ can distinguish malicious recommender agents that may lie or give inconsistent information. 

\begin{equation*}
rt(b,c,rtd)
\end{equation*}

An agent uses two different data structures to store trust information, namely set $Q$ and $R$. Also, the outcome of an experience $e$ is graded with one of the four degrees in set $E = \{vg,g,b,vb\}$ respectively \textit{very good (vg), good (g), bad (b), very bad (vb)}. They correspond to the four trust degrees above. 

$Q$ is for direct experiences and it holds agents' name (set $A$), experience results (set $S$) and contexts of trust (set $C$). In set $S$, there are four values $s=(s_{vg}, s_{g}, s_{b}, s_{vb})$ for experiences with each agent, so $S=\{(s_{vg}, s_{g}, s_{b}, s_{vb})\}$. After each experience the corresponding value incerements by one. For instance, we have in the beginning the set (0,0,0,0)  respectively $(s_{vg}, s_{g}, s_{b}, s_{vb})$. If the result of the experience is \textit{good (g)}, the set will be (0,1,0,0). This feature of trust is called $dynamism$ and it means that trust can be changed at any time.

\begin{equation*}
Q \subseteq C \times A \times S
\end{equation*}

$R$ holds information of recommender agents. The purpose of this set is to adjust the recommendation according to our perception. A parameter called \textit{semantic distance} helps us to measure the similarity between the opinions. We need this parameter because according to the trust characteristics, trust is a “subjective probability” and not transitive.  For example, an agent $x$ recommends that agent $y$ is $untrustworthy$. However, our opinion is the agent $y$ is $trustworthy$. We should adjust the future recommendations from $x$ by incrementing the trust value one (because there is only one grade between $trustworthy$ and $untrustworthy$). This approach makes the model robust against malicious recommenders. 
\begin{equation*}
R \subseteq C \times A \times T
\end{equation*}
  
For every recommendation, we store recommenders' name (set $A$), contexts (set $C$) and the adjustment set.  For adjustments, we have the set $T$ which consist of four other sets $\{T_{vg}, T_{g}, T_{b}, T_{vb}\}$. After an adjustment is made, the value is stored in the corresponding set. The value of an adjustment is the difference between grades. If we take the example above the set $T$ will be $(\{ \}, \{ \}, \{1\}, \{ \})$.

The first formula used in the model is for direct trust. As mentioned above, after every interaction with an agent the corresponding value in set $S$ is incremented by one (1). The maximum of these values is accepted as the trust degree (2). If there is more than one value for $max(s)$, the table introduced in the article for uncertain values is used.

\begin{equation}
e \in E, s_e = s_e + 1
\end{equation}
\begin{equation}
\exists td \in E \quad \forall s_e \in s, \:(s_e=max(s)) \implies (td = e)
\end{equation}

To evaluate the trustworthiness of recommenders, all adjusment values in set T are used. Let $T^a = T_{vg} \cup T_{g} \cup T_{b} \cup T_{vb}$. To get the \textit{recommender trust degree (rtd)}, we take the $mod$ of the absolute values in $T^a$ (3).

\begin{equation}
rtd = mod (\{ \forall x\in T^a \mid |x|\})
\end{equation}

For each type of experience $e \in E$ there is a \textit{semantic distance} value in the tuple $sd = (sd_{vg}, sd_{g}, sd_{b}, sd_{vb})$. The value is calculated by taking the mod of the corresponding $T$ set (4).

\begin{equation}
\forall e \in E, \: sd_e = mod(T_e)
\end{equation}

Evaluating a recommendation is done by combining the \textit{recommendation (rd)} with the corresponding \textit{semantic distance} $sd_{rd}$ (5). For example, if the \textit{recommendation (rd)} is \textit{good (g)}, we combine the semantic distance for $good$ $(sd_{g})$ with $rd$. The new recommendation value is denoted by $rd^*$. The symbol $\oplus$ denotes a simple summing operation over the trust degrees. Let assume $sd_g = 1$ and $rd$ is $good$, the result will be $g \oplus 1 = vg$.

\begin{equation}
rd^* = rd \oplus sd_{rd}
\end{equation}
 
If an agent gets multiple recommendations, they are combined using \textit{recommender weights}. These weights are used to calculate the particular effect of the recommender on the recommendation result. The $rtd$ value of the recommender determines its \textit{recommender weight}. The possible $rtd$ values $(0,1,2,3)$ are matched to the weights $(9,5,3,1)$ respectively (i.e. an agent with $rtd = 0$ has the weight $9$). Additionally, if the $rtd$ value is unknown, the weight is $0$. 

To combine recommendations the weights of all recommenders who recommend an agent $a$ for a certain experience $e \in  E$ are added up (6). $L_a$ is the set of all weight values of the recommenders of an agent $a$. 

\begin{equation}
\forall e \in E \quad \forall w_i \in L_a, sum_e = \sum_{i=1}^{|L_a|} w_i  
\end{equation}

The maximum of $sum_e$ values is the \textit{combined trust degree (ct)}. As it is the case for trust degrees, if we have more than one maximum $sum_e$, the uncertainty table should be used to calculate the result. 

Finally, the authors recommend that some information should exist in the knowledge base of a newcomer agent before it joins a community. For example, there should be some recommendations in set $R$ before the agent starts interacting. Because the authors think that in the real-life a newcomer gets some information about an environment before joining it.

\subsubsection{Applicability}
The model is well defined and the characteristics of trust in social life are clearly represented. The mathematical formulas are easy to understand and implement, which is one of the goals of the article. An example application is also provided to show the applicability. However, this example is only a mathematical demonstration. The model was not implemented and tested in a virtual medium. So the real performance of the overall system is unknown to the readers of the article.

The only problem in the model which is mentioned in \cite{b9} is that the system cannot distinguish between a liar agent and an agent that thinks differently. An agent tries to interpret a recommendation according to its own perception. For example, a non-malicious recommender who just thinks differently says to another agent what he thinks. But if the other agent has a different perception, he will mark the recommender as untrustable for future recommendations. In this case, it is not possible to decide if the recommender is a malicious agent or just has a different opinion. This may not seem important in general but some particular cases should also be taken into account for the robustness of the system. For example, if these two types of agents are equated and not distinguished, the agent that thinks different than us cannot give recommendations in the future, because he/she is marked as untrustworthy. So we will lose an agent who is maybe trustworthy but has a different opinion.

\subsection{Comparison}
Here, the reference model is compared to the base model.

The trust and reputation model of the base article is a rather self-enclosed system. This means evaluations of the trust degree is done within the system of an agent and the result matters only to the agent itself. Other participants in the community do not know how their behavior is evaluated by a counterparty. In contrast with this, the model of the reference article uses incentives and sanctions. For example, agent $a$ says to agent $b$ that he will get a 0.7 trust point when he performs a certain action. So agent $b$ has a motivation to get an incentive from agent $a$. Also, a sanction system is used for the exact opposite of this example. The incentive and sanction system makes agents more careful in interactions.

Another important difference in the models is the forgiveness notion. We know this notion from our social life. We forget other people’s mistakes over time and forgive them. This concept is excellently implemented in the reference system. New values have a stronger influence in the calculation than previous values. This helps agents who change their behavior to integrate into the system more easily \cite{b3}. A forgiveness system similar to this is not implemented in the base article.

Moreover, the propositions to put these models into practice are different in the articles. In the base article, the system performance is not taken into account. The base article provides only the mathematical implementation of the model and does not consider performance parameters of the system e.g. storage. The main focus of the article is to provide a social point of view for defining a trust model rather than creating an efficient system. On the other hand, the reference article aims to create a powerful model that uses system resources, such as storage, efficiently.

\section{The Trust Model}
In this section, we will provide our approach for a trust model. We have analyzed the base and the reference article sufficiently to develop this model. The reference article was especially important when deciding on the trust metric and the model parameters. However, our model has a very different perspective than both articles in general. Also, we have added another aspect, which does not exist in neither articles, the risk value.

\subsection{Explanation of the model}
Our model acts mainly based on user preferences. The user places the agent within the system and sets the threshold values as he/she wishes, which are introduced below in section $G$. The algorithm in section $H$ decides on the two characteristics of an agent with the help of threshold values. These characteristics are “trustworthiness” and “riskiness”. The user should decide what kind of agent he wants to work with. For example, if the user wants to work with a trustable agent and the riskiness is not important to him, he should adjust the settings of the system accordingly.
\subsection{Trust metric}
Various metrics are used in the previous work to measure trust. For example, the binary metric measures trust with two degrees (-1,1) namely good and bad, or the continuous metric has continuous values to measure trust in a certain range as in [-1,1]. Some of these metrics are listed and assessed according to their advantages and disadvantages in the reference article\cite{b3}. For our model, we have decided on a discrete trust metric in the range [0,1] which has a step size of \textit{0.1} (Figure I). After each interaction, we choose a value from the range according to the agent's behavior.

\begin{figure}[htbp]
\centering
\caption{The range of the trust metric}
\begin{tikzpicture}[scale=0.7]
\centering
\filldraw (1,0) circle[radius=2pt]
          (11,0) circle[radius=2pt];
\foreach [count=\i] \x  in {0.0,0.1,0.2,0.3,0.4,0.5,0.6,0.7,0.8,0.9,1.0}
   \draw (\i,1pt) -- (\i,-2pt) node[below,fill=white] {\x};
\draw[-] (1,-1)--++(-90:.3) node[below]{Untrustworthy};
\draw[-] (11,-1)--++(-90:.3) node[below]{Trustworthy};
\draw[thick] (1,0) -- (11,0);
\end{tikzpicture}
\label{fig:my_label}
\end{figure}

The user himself should define the exact meaning of the values in this range according to behavior types. For example, a calculation request to an agent can be resulted in the trust degree \textit{0.7} which might correspond to ``the request was replied and the answer was correct, but it was delayed''. Or as another example, the trust degree \textit{0.2} can correspond to the behavior ``the request was replied but the answer was wrong''. So the user can define these values as he/she wishes. We will not implement this part in our algorithm and we expect the user will define them.

\subsection{Experience Set}
The trust degrees resulting from our interactions with other agents are stored in a \textit{experience set}. When the set is filled with values, it is called an \textit{interaction period}. The size $s$ of this set starts with 1 and increases by one after every \textit{interaction period} up to a certain size $n$. For example, in the first interaction $s=1$, in the second and third interactions $s=2$ and so on. The limit $n$ of this increment is determined by the user beforehand. We have termed each trust degree in the experience set with the variable $TD_i$, for $0\leq i < s$, where \textit{i} is the interaction number in the experience set. An example experience set is shown in Table I where $s=8$. After every \textit{interaction period}, the \textit{general trust degree} $TD_{gen}$ will be calculated from the experience set and then the set will be reset.
					
\begin{table}[htbp]
\centering
\caption{An example experience set}
\begin{tabular}{|l|l|l|l|l|l|l|l|}
\hline
\textit{$TD_0$}& \textit{$TD_1$} & \textit{$TD_2$} & \textit{$TD_3$} & \textit{$TD_4$} & \textit{$TD_5$} & \textit{$TD_6$} & \textit{$TD_7$} \\ \hline
 0.8 & 0.7 & 0.8 & 0.6 & 0.7  & 0.5 & 0.2  & 0.7  \\ \hline
\end{tabular}
\label{tab:my_label}
\end{table}

\subsection{Calculating General Trust Degree}
The general trust degree $TD_{gen}$ holds the trustworthiness of an agent which is calculated using (I) the experience set and (II) the latest $TD_{gen}$. In our approach, the opinion about the trustworthiness of an agent is less updated as the “degree of acquaintance” increases. $TD_{gen}$ is updated after every \textit{interaction period}. The number of interactions in the \textit{interaction period} increases by 1 after the period is completed. It means the better we get to know an agent, the less the trust degree is updated. This provides us to behave at the beginning of communication more carefully. However, when we have more experience with an agent, our opinion about him/her becomes more persistent. The limit for this update frequency can be adjusted with the number $n$ from the user beforehand.

We will first get an average value from the experience set for calculation. For that, we have decided to use the median value instead of the mean value because the mean value is highly sensitive to outlier data. For example, we have different values in a set which are distributed around ‘0.2’. A new value of ‘0.8’ will change the mean value too much. However, the median value is not sensitive to outlier data like mean value and it uses the middle value of an ordered set (7). We have defined the variable $EX_{med}$ which holds the median value of the experience set. This variable is calculated after every \textit{interaction period} again.
\begin{equation}
EX_{med}= \begin{cases}
TD_{(s+1)/2}, & \text{when \textit{s} is odd}  \\
(TD_{s/2} + TD_{(s+1)/2})/2, & \text{when \textit{s} is even}
\end{cases}
\end{equation}

We ignore extreme values in an experience set by using median because we want to calculate the trust degree according to the general behaviors of an agent. But it does not mean that we ignore these extreme behaviors of the agent completely. We will take them into account with  \textit{risk value} which we will introduce below.

The second required parameter to calculate the new $TD_{gen}$ is the latest $TD_{gen}$. To avoid confusion we have attached to these variables a number \textit{m}, where $0\leq \textit{m}$. So the new general trust degree will be $TD_{gen\_m}$ and the latest general trust degree $TD_{gen\_(m-1)}$. We take into account past experiences by considering the latest trust degree. However, past experiences should not have the same effect as the current experiences. As mentioned in \cite{b3}, current experiences are more important than past experiences. Therefore the effect of the latest trust degree depends on the parameter \textit{k}, which $0 < \textit{k}$. The higher \textit{k} the lower effect of the past experiences in our calculation (8). This provides us also the social notion of ‘forgiveness’. So if an agent changes its behavior, we forgive it by reducing the effect of the past behaviors in the calculation. 
 \begin{equation}
TD_{gen\_m} = (TD_{gen\_(m-1)} + \textit{k} \cdot EX_{med}) / (\textit{k} + 1)
 \end{equation}
 
If there is no previous $TD_{gen}$ to calculate a new one, then $TD_{gen} = R_{gen}$ where $R_{gen}$ is the \textit{general reputation value} (see below).

\subsection{Risk Value}
Another important parameter in our model is the risk value $RV$. This parameter indicates how risky it is to contact an agent. If an agent has very unstable behaviors, it will be evaluated as risky. If its behaviors are not complex, it is not a risky agent. So the variation of the values in the experience set expresses the agent’s riskiness.  To calculate this variation we use semi-deviation which is also used in other fields such as finance to calculate risk. As distinct from standard deviation, semi-deviation measures only the negative fluctuations. In our case, it is a better option to use it because we should accept an agent as risky if his/her behaviors change in the negative direction. The changes in the positive direction do not indicate the riskiness. For this calculation, we need the mean value of the experience set (9). The values lower than the average value should be taken into account (10). The semi-deviation of the experience set is our risk value $RV$ (11). 
			
\begin{equation}
x = \frac {\sum_{i=0}^{\textit{s}}TD_i}{s} 
 \end{equation}
\begin{equation}
f(y) = \begin{cases} 1, & \text{when  $y < x$ } \\
0 , & \text{when  $y \geq x$ } 
\end{cases}
\end{equation}
\begin{equation}
RV = \sqrt{\frac{\sum_{i=0}^{\textit{s}}((TD_i - x)^2 \cdot f(TD_i))}{N}}
\end{equation}
\begin{equation*}
\text{$N=$ \textit{the number of values lower than the average value}}
\end{equation*}

If we contact with an agent for the first time, then the risk value $RV$ will be equal to the threshold $RV_{th}$. Otherwise the agent will be evaluated as risky and this evaluation will be biased.

\subsection{Reputation value}
Since we do not have a general trust degree $TD_{gen}$ at the beginning of communication with an agent, we need the reputation values at this point. But for later calculations, we will not use them. In other words, we use reputation values as an aid to start a communication.

We calculate the general reputation value $R_{gen}$ with a very similar way to the median of the experience set $EX_{med}$. Firstly, we get $n$ recommendations from other agents and each recommendation is represented with the variable $R_i$ where $0 \leq i < s$. Afterward, we sort recommendation values and then get their median (12). We accept this median value as the $TD_{gen\_0}$.

If we cannot get any reputation value from others, the general trust degree $TD_{gen}$ will be equal to the threshold value $TD_{th}$, which will be introduced in the next section. Otherwise, the agent will be evaluated as untrustworthy in the beginning and this opinion will be biased.
\begin{equation}
R_{gen} = \begin{cases}
R_{(s+1)/2}, & \text{when \textit{s} is odd}  \\
(R_{s/2} + R_{(s+1)/2})/2, & \text{when \textit{s} is even}
\end{cases}
\end{equation}
\subsection{Thresholds}
Our model is flexible during the decision-making process for evaluating trust. We do not have default thresholds for parameters. The threshold values should be given by the system administrator who set up the agent. For example, if the system owner sets the threshold of the trust degree to 0.4, we will accept a value over 0.4 as trustworthy and any values below this threshold will be considered as untrustworthy. The same is valid for risk value. If the risk value exceeds the given threshold, we then accept the agent as risky. The thresholds are named as the following:
\begin{itemize}
	\item $TD_{th} \rightarrow$ Threshold for the trust degree
	\item $RV_{th} \rightarrow$ Threshold for the risk value
\end{itemize}
\subsection{Algorithm}
We have explained the details of the trust model so far. In this section, we will provide an algorithm that uses the parameters introduced in the previous sections. It calculates the final result for the agent to decide whether to contact the other party or not.

\begin{algorithm}[H]
	\caption{Algorithm for the trust model}
	\begin{algorithmic}[1]
	    \State a = getAnAgent()
	    \If	{firstInteraction}
	    \State $RV = RV_{th}$
	    \If{isThereAnyRecommandationFor(a)}
	    \State $TD_{gen}$ =  $R_{gen}$
	    \Else
	     \State $TD_{gen}$ = $TD_{th}$
	    \EndIf
	    \Else
	     \State $TD_{gen}$ = calculateTD\_gen() \Comment{\textit{Eq.: 8}}
	     \State $RV$ = calculateRV() \Comment{\textit{Eq.: 11}} 
        \EndIf
        
        \If{($TD_{gen} \geq TD_{th}$) and ($RV < RV_{th}$)}
        \State trustworthy = True
        \State risky = False
        \ElsIf{($TD_{gen} \geq TD_{th}$) and ($RV \geq RV_{th}$)}
        \State trustworthy = True
        \State risky = True
        \ElsIf{($TD_{gen} < TD_{th}$) and ($RV < RV_{th}$)}
        \State trustworthy = False
        \State risky = False
        \ElsIf{($TD_{gen} < TD_{th}$) and ($RV \geq RV_{th}$)}
        \State trustworthy = False
        \State risky = True
        \EndIf
        \State characteristics = [trustworthy, risky]
        \State \textbf{return} characteristics

	\end{algorithmic} 
\end{algorithm}

\section{Conclusion}
We cannot imagine a social life without the concepts of trust and reputation. They have also a significant effect in other areas such as computer science. Therefore, simulating these concepts in a virtual medium is an interesting research area. In this paper, we provide a mathematical implementation of these concepts to apply them in a virtual medium. Our goal was to provide an approach that considers mainly the social characteristics of the concepts. To do this, we have first given the definitions of trust and reputation in a social sense. Then we have reviewed two articles in this research area and compared them. Afterward, we have introduced our trust model, which is based on our definitions. The first essential feature of our model is flexibility. The thresholds in the model can be changed by the user, hence the behavior of the model is changeable according to the user's wishes. The second notable feature is the implementation of social notions, such as forgiveness.  

As future work, an implementation of the proposed model could provide interesting results.

\end{document}